\documentclass[10pt,letterpaper]{article}
\usepackage{hyperref}
\usepackage{cogsci}

\cogscifinalcopy % Uncomment this line for the final submission 

\usepackage{pslatex}
\usepackage{apacite}
\usepackage{float} % Roger Levy added this and changed figure/table
\usepackage{graphicx}
\usepackage{lipsum}% http://ctan.org/pkg/lipsum

\title{Shapeception: Unravelling Brain Activity during Animated Shape Perception and Mentalization
}
 
\author{{\large \bf Varad Srivastava (varadsrivastava.iitdelhi@gmail.com)} \\
  Indian Institute of Technology Delhi \\
  New Delhi, India
  \AND {\large \bf Minaxi Goel (minaxi.goel@research.iiit.ac.in)} \\
   International Institute of Information Technology Hyderabad \\
  Telangana, India}

\begin{document}

\maketitle

\begin{abstract}
% Relationship between mentalizing and empathy have been studied using human features as stimuli. Here, we investigate this relationship by incorporating animated shapes, that evokes mental state attributions. fMRI analysis revealed that human could empathize while mentalizing interacting shapes, despite lacking human features like facial expressions or body postures. We also observed a region close to the pSTS, which is less activated in autistic individuals, suggesting that our approach can help diagnose disorders linked with impairment of the ability to mentalize and empathize. Besides, our machine learning model was also able to model the prediction of mentalization, using fMRI data, with promising results. Moreover, it also made reliable predictions when trained on the sole ROIs, which were observed to be pronounced during the mentalizing task. Therefore, a similar predictive modelling approach could be performed for the diagnosis of such social cognition impaired disorders and for the assessment of the mentalization-based therapies.

In this paper, we investigate the brain activity elicited during perception of animated shapes as stimuli, which have been found to evoke mental state attributions. Contrary to a previous study, we incorporated the participants' responses in the analysis, and observed robust activations in mPFC, which has been found to play an important role in understanding other’s and one’s own nature. From our analyses, TPOj was observed showing robust activation during the task as well as functionally connected to AA and LTC, which lead to speculation that empathy might co-occur with mentalizing in the task and that humans might be able to empathize with these interacting shapes, in spite of lacking human features. Along with this, in one of our analyses, we were able to localize a region close to the pSTS, where the activation depicted the participants' 'ability to mentalize'.
Based on our observations, we modelled the prediction of mentalization and propose our model as an approach towards developing a brain-activity based model to detect ToM (Theory of Mind) difficulties, which could be useful in research about disorders like Autism Spectrum Disorder (ASD) as well as assessment of mentalization-based
treatments. Additionally, we use our findings to reiterate how the Resting State might not always act as a good control condition and that control conditions should be task-specific.

\end{abstract}

\section{Introduction}
During its developmental stages, the brain forms memories and uses these as templates to infer others' emotional and mental states. Interpreting others' experiences and behaviors also shapes one's own behavior. This process of ascribing mental states to others, inferring their emotions and feelings, and reflecting on them is called mentalizing. The ability or the capacity to make these inferences and dissociate others' mental states (thoughts and feelings) to gain an understanding of their mind is called Theory of Mind (ToM). This understanding is used to explain and predict behaviour of other people. ToM plays a critical role in development of sense of self \cite{fonagy1991thinking, fonagy1996playing}. 

It is central to cognition and emotion, and fosters understanding of emotional states of others as well as our own, that further guide our behavior. The concept of mentalization was first conceptualised by \citeA{fonagy2012mentalization} of Anna Freud Centre. They had defined mentalizing as ``being engaged in a form of (mostly preconscious) imaginative mental activity that enables us to perceive and interpret human behavior in terms of intentional mental states", such as needs, desires, thoughts, feelings, beliefs, attitudes, intentions, and reasons \cite{allen2008mentalizing}. 

It also enables a person to make out what other person is going through by analyzing their behavior such as body postures, color of speech, words, facial expressions and eye-movements. For e.g., if someone approaches us with a dull body language, we are able to figure out that something is wrong with the person, or we can make out if someone is interested or not in our conversation by observing the other person's eye-movements. All these cues help in deducing other person's mental states that further enables empathizing with them. 

As part of the Human Connectome Project in the Theory of Mind (ToM) domain, 339 participants were presented with animated videos of shapes (circles, triangles, or squares) either interacting with each other or moving randomly on the screen, and the task-evoked functional brain activity data was recorded. These interacting shapes have previously shown evidence for mental state attributions \cite{castelli2000movement,abell2000triangles,white2011developing}. In this study, we perform analyses of this brain activity during mentalisation and observe some significant findings.

In one of our analyses, we observed that TPOj showed robust activation
during the task. Further, from functional connectivity analysis, we established that TPOj is functionally connected with AA and LTC, while mentalizing. As TPOj is associated with empathy, we speculate mentalizing and empathy might co-occur even in this case, as previous research posits \cite{cerniglia2019intersections, schnell2011functional}.

% A study \cite{decety2013fmri} did research on psychopaths, where they (psychopaths) are defined as lacking in empathy and remorse. When they were asked to imagine themselves in pain, normal activity in brain regions involved in empathy for pain were observed, such as anterior insula, the anterior midcingulate cortex, somatosensory cortex, and the right amygdala. But, when they were asked to imagine for others, these areas became inactive. 

% In stressful situations, people with BPD tend to ‘lose’ mentalizing, and in doing so make others ‘lose’ mentalizing.

We noted that for our study, a task-related condition is a better control than the Resting State (RS), hence, reaffirming that control conditions should be task-specific and that RS might not always act as a perfect control condition. Additionally, we obtained robust activations in a region close to the pSTS, where the activations depicted the participants' 'ability to mentalize'. Since this region has been linked to disorders with abnormality in ToM like Autism Spectrum Disorder (ASD), this finding holds immense significance.  

Finally, we also perform dimensionality reduction to look out for meaningful and interesting patterns within this high dimensional brain activity data \cite{bernstein_varad}. We found that the projections of the brain activity in lower dimensions during various conditions (for e.g. mentalizing and not mentalizing) get structured into distinct clusters. In
light of this observation, we modelled the prediction of 
mentalization. The model was able to generate distinct decision boundaries and returned impressive  accuracy in predicting about mentalization. Along with our findings and observations of the brain activity during perception of these animations, we propose our model as an approach towards developing a brain-activity based model
to detect ToM (Theory of Mind) difficulties, which could be useful in research about disorders like Autism Spectrum Disorder (ASD) as well as for assessment of mentalization-based treatments.

\section{Method}
\subsection{Data}
We used the task-evoked functional brain activity data from the Social Cognition (Theory of Mind) domain of the Human Connectome Project. 339 participants were presented with animated videos of shapes (circles, triangles, or squares) either interacting with each other or moving randomly on the screen. These videos were developed by either \citeA{castelli2000movement} or \citeA{wheatley2007understanding}. For e.g. in the interaction one, two shapes (a big triangle and a small one) were animated to imply complex mental states, like being involved in persuading, bluffing, mocking, surprising one another or even depicting an intention to deceive. In random movement, the shapes were showed to be bouncing off the walls or just drifting about.
% In this study, stimuli were projected onto a computer screen behind the subject’s head within the imaging chamber and the screen was viewed by a mirror positioned approximately 8 cm above the subject’s face. 

% An example can be seen in FIGURE.
% --- kinds of shapes/stimuli showed image ---

After presentation of each video, participants were asked to judge whether they perceived the shapes as having `mental interaction' (an interaction in which shapes appear to be understanding each other's mental states i.e. feelings, thoughts or intentions), `random movement' (i.e. the movement of shapes appears to be random without any kind of interaction), or `not sure'. Since, the perceptions of the participants can be directly attributed to their ability to reflect on the emotional and instinctive responses in the shapes which are presented, hence, their responses reflected their ability to mentalize.

% For recording the fMRI activity, each subject’s Blood Oxygen Level Dependent (BOLD) signals were collected for both the resting state and task evoked active state. BOLD signals are observed due to the difference in magnetic properties of oxygenated and deoxygenated blood, which is captured by the MRI machine. 
 
\begin{figure}
\begin{center}
\includegraphics[width=3in]{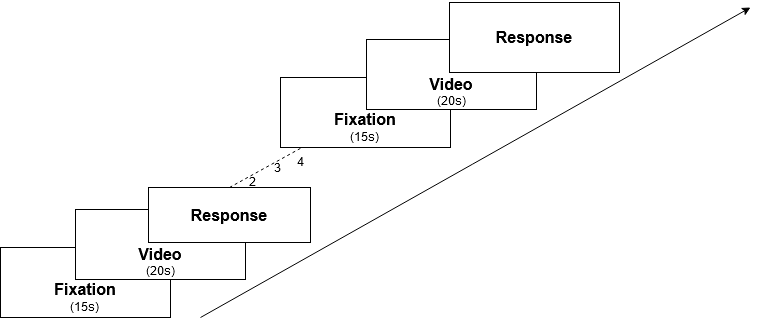}
\end{center}
\caption{Experiment Design} 
\label{expdesign}
\end{figure}

In the task-based fMRI, the data was collected during two runs. Each run consisted of 5 video blocks with mental interaction and random movements that were shown to participants and their responses for each video were recorded. Each video had a fixation block of 15 seconds and they lasted 20 seconds. Each run had either 2 videos of mental interaction and 3 videos of random movement or vice-versa. The experiment design is demonstrated in Figure~\ref{expdesign}. The dataset used in this work is curated and made available by the Neuromatch Academy\footnote{https://osf.io/hygbm/}.

\subsection{Conditions with Nomenclature}
\begin{figure}
\begin{center}
\includegraphics[width=5cm]{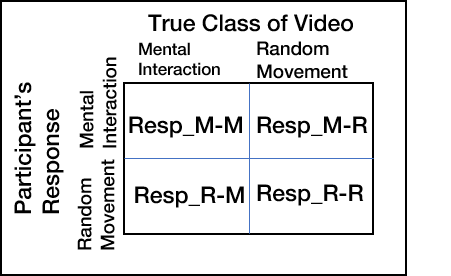}
\end{center}
\caption{Nomenclature followed while performing analysis} 
\label{nome}
\end{figure}

This is the nomenclature that we follow for the responses of participants and the conditions of videos:\\
Resp\_X-Y denotes the response belonging to ‘X’ category [M: ‘Mental Interaction’, R: ‘Random Movement’], and Y denotes the condition of videos [M: ‘Mental’, R: ‘Random’] (Figure~\ref{nome}). We follow this nomenclature hereon.

\subsection{Analysis}
We performed four kinds of analyses for each of the following objectives. First, identifying the parcels, regions and networks activated, as shown in '\nameref{method_1}' section. Second, investigating the functional connectivity among these regions of interest (ROIs), as discussed in '\nameref{method_2}' section. Third, dimensionality reduction to look for meaningful patterns, as discussed in '\nameref{method_3}' section. And fourth, modelling the prediction of mentalization, as discussed in '\nameref{method_4}' section. The code for these analyses is available on Github\footnote{https://github.com/varadsrivastava/shapeception}.

\subsubsection{Identifying regions of activity}
\label{method_1}
We performed subtraction analysis to identify the regions and networks that were activated while mentalizing. The parcels were identified using parcellations by \citeA{glasser2016multi}, one of the most comprehensive human cerebral cortex parcellations, where 180 symmetric areas per hemisphere have been delineated as parcels. These 180 parcels were also grouped into 22 regions based on geographically contiguous areas which can be observed in their entirety when viewed from a single perspective and on the basis of their common architectures, task-fMRI profiles, or functional connectivity. 

Understanding cognitive abilities requires studying and characterizing the architecture of human brain across multiple levels of organization. Therefore to investigate activity at the level of networks, we also use network assignments of parcels from \citeA{ji2019mapping}, which provides a comprehensive whole-brain functional network atlas. 
Using Pearson correlation, each voxel was assigned to the network with which it shared the highest mean connectivity across the parcels.
Additionally, parcellations on the fsaverage5 surface \cite{mills2016hcp} and approximate MNI coordinates of each region were used for the visualizations.

Hence for each participant, fMRI BOLD time series were extracted from the 360 independently identified parcels and subtraction analysis was performed for the following combinations of conditions and responses:
\\
\\
Case A: Mental and Random condition of Videos\\
Case B: Resp\_M-M and Resp\_R-R (taking Resp\_R-R as control)\\
Case C: Resp\_M-M and Resting State (RS) (taking RS as control)\\
Case D: Resp\_M-M and Resp\_R-M\\

% \begin{enumerate}
% \item
% Case A: Mental and Random condition of Videos
% \item
% Case B: Resp\_M-M and Resp\_R-R (taking Resp\_R-R as control)
% \item
% Case C: Resp\_M-M and Resting State (taking resting state as control)
% \item
% Case D: Resp\_M-M and Resp\_R-M
% \end{enumerate}

\subsubsection{Functional Connectivity}
\label{method_2}
Correlational analysis was performed to investigate the functional connectivity among the ROIs (as discussed in \nameref{roi} section). For each participant, fMRI BOLD time series were extracted from the 36 identified parcels in which robust activations were observed, corresponding to the Resp\_M-M condition in each run. Subsequently, Pearson product-moment correlation coefficients between each pair of parcels were calculated for each participant. A group average functional connectivity matrix was formed by averaging across all participants for each parcel. A functional connectivity matrix for N parcels is defined as a N×N matrix M, where M(i, j) contains the Pearson correlation coefficients between parcels i and j. Hence in this way, a 36 × 36 functional connectivity matrix was obtained and plotted with parcels arranged region-wise, as assigned by \citeA{glasser2016multi}.

\subsubsection{Dimensionality Reduction}
\label{method_3}
To observe this high-dimensional data, we also performed dimensionality reduction using t-Distributed Stochastic Neighbor Embedding (t-SNE). It is a non-linear dimensionality reduction algorithm used for exploring high-dimensional data, which finds patterns in the data based on similarity of data points with multiple features.
fMRI BOLD time series (for parcels from the four ROIs) were extracted for conditions Resp\_M-M and Resp\_R-R (i.e. when shapes in the `Mental' and `Random' condition of videos were correctly perceived as having mental interaction and random movement, respectively), Resp\_R-M (i.e. when shapes in the 'Mental condition' were incorrectly perceived), as well as the Resting State. These were then averaged across time frames and subsequently concatenated with each other, hence resulting in a 1356 x 36 NumPy array. Using t-SNE, we mapped these high-dimensional time-series to two dimensions, where each point represents a participant's scan.

\subsubsection{Predicting Mentalization}
\label{method_4}
In the corresponding lower two-dimensional space, we observe three distinct clusters that could be identified as belonging to three conditions: Resp\_M-M, Resp\_R-R and Resting State (we exclude the Resp\_R-M condition, as the mappings for Resp\_M-M and Resp\_R-M conditions do not seem to be segmented well).
Therefore, we further investigate if mentalization could be predicted based on this two-dimensional mapping of the brain activity in the four ROIs. We use the k-nearest neighbours classifier for this purpose.

The distinct clusters correspond to the three classes: `0' for Resp\_R-R, `1' for Resp\_M-M, and '2' for the Resting State. To chose the optimal value of 'k' (no. of nearest neighbours) for the k-nearest neighbours classifier, we use the Elbow method.
When 'k' is small, we limit the range of a given prediction and force our classifier to be ignorant of the overall distribution (more complex model and may lead to overfitting). A small value for 'k' yields the most adaptable fit, with low bias but high variance. This would result in the decision boundary being more jagged.
A higher 'k', on the other hand, averages more points in each prediction and is thus more resistant to outliers. As expected, these would then result in smoother decision boundaries (less complex model), decreasing variance but increasing the bias.

\begin{figure}
\begin{center}
\includegraphics[width=8cm]{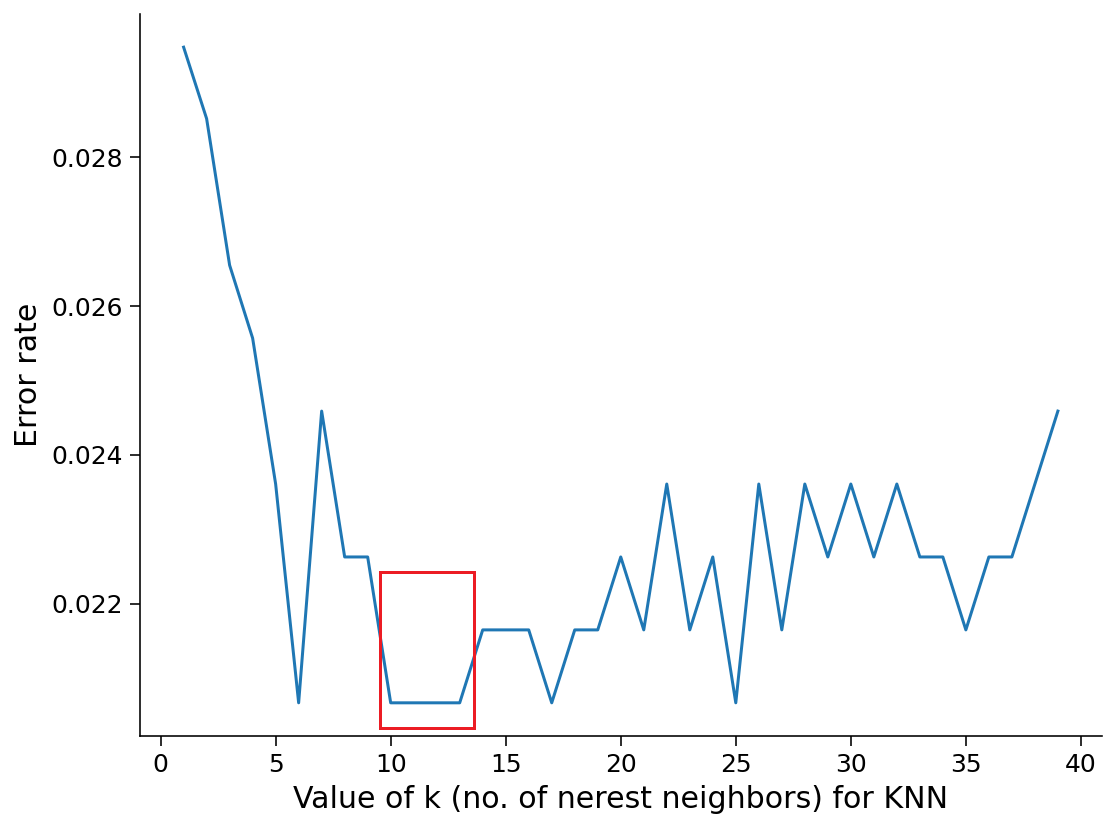}
\end{center}
\caption{Using Elbow method to determine optimal value of k for KNN classifier} 
\label{elbow}
\end{figure}

Models that are either too complex or not complex enough are penalised in the form of error rate. When the model has the appropriate level of complexity, we get the lowest error rate. We performed a 10-fold cross validation to estimate the error rates for various values of 'k'. As seen in Figure~\ref{elbow}, in our case, the error rate is the lowest around 9\textless k \textless 14. Hence, we select k=11 as the hyper-parameter.

In this way, we investigated whether a lower dimension mapping of the activity in the 36 parcels belonging to four ROIs (\nameref{roi} section) where we found the most robust activity could be used to distinctively predict mentalization.

We also checked for overfitting by evaluating the model on test data that it has not experienced before, by splitting the data 70-30 into training and testing sets, after shuffling it.
% In this approach, k-fold cross-validation process is replicated multiple times and mean performance of the model across all folds (of the same dataset) and repeats is reported. RepeatedKFold class from the Python scikit-learn library is used for the implementation. Two main parameters were set: n\_splits (i.e. number of folds) = 10, n\_repeats (i.e. number of repeats) = 3.

% \begin{table}
% \begin{center} 
% \caption{Parameters} 
% \label{parameters} 
% \vskip 0.12in
% \begin{tabular}{ll} 
% \hline
% Parameter    &  Value \\
% \hline
% Maximum iterations        &   300 \\
% Fit intercept   &   True \\
% Penalty           &   L2 \\
% Solver           &   lbfgs \\
% \hline
% \end{tabular} 
% \end{center} 
% \end{table}

\begin{figure*}
  \includegraphics[width=\textwidth]{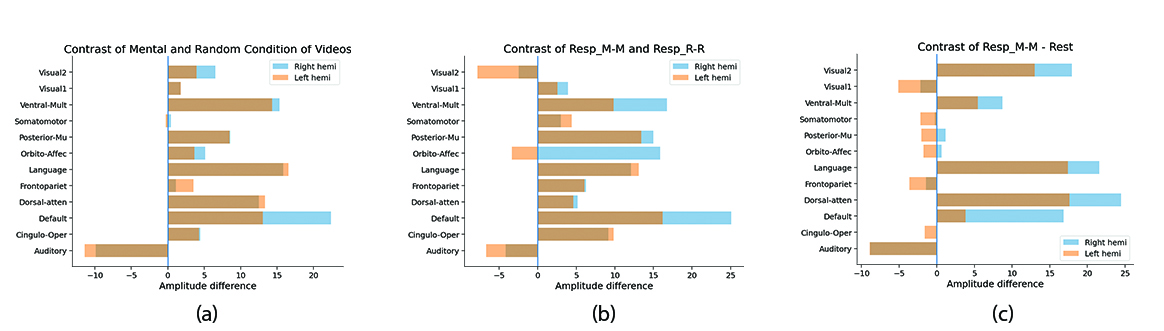}
  \caption{Activations in networks during mentalizing for the three cases: (a) Case A, (b) Case B, and (c) Case C}
  \label{result1}
\end{figure*}

\section{Results}
\label{roi}
\subsection{Identifying regions of activity}
\begin{figure}[h!]
\begin{center}
\includegraphics[height=3cm,width=5cm]{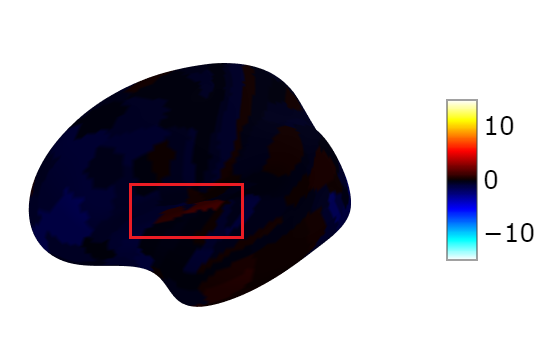}
\end{center}
\caption{Activations in Case D (lateral view)} 
\label{result4}
\end{figure}

Subtraction analyses yielded robust activations across temporo-parieto-occipital junction (TPOj), medial prefrontal cortex (mPFC), auditory association (AA) area, and lateral temporal cortex (LTC). These findings are consistent with previous research that have observed regions associated with mentalizing as mPFC, posterior superior temporal sulcus (pSTS), temporoparietal junction (TPj) and temporal poles (TP) \cite{silani2013right, frith2001mind, cerniglia2019intersections, hooker2008mentalizing, castelli2000movement}. 
% and the orbital/ventral region of the PFC, i.e. the orbito-frontal cortex (OFC) is involved in experiencing the shared affective experience during cognitive mentalizing \cite{cerniglia2019intersections}.
% Robust activations were observed in TPOj, which has been associated with empathy \cite{coll2017effect, mai2016using}. This suggests that humans do empathize with interacting shapes which elicit mental state attributions, and hence, kinetic properties are enough to evoke empathy.
Four interesting observations are:
\begin{enumerate}
\item
Default mode network (DMN) was found to be consistently activated with high contrasts across Cases (A,B,C) that we investigated (Figure~\ref{result1}).

\item
 We observed more promimnent activation in the mPFC for the mentalizing task in Case B (Figure~\ref{result2}b), when participants' responses are taken into account, compared to Case A, where they are not.
 
%  We also did not see robust mPFC activation in the social cognition task,
% which would have been expected based on prior studies. In this case,
% tSNR was not particularly low in the more dorsal part of medial PFC,
% though it was lower in subgenual regions. Thus, SNR may not be the
% sole explanation for the lack of activation in this region in the currently analyzed dataset (n = 20). Alternative analyses that might reveal
% activation in medial PFC during the social cognition include individual
% difference approaches, or analyses that code trials as a function of the
% participant's evaluation of the film clip.

% We observed more clear contrast for the regions involved in empathy when the participant responses were considered in our study, as compared to when they were not considered, that is only condition-based analysis (Case A) was performed as in the study done by \cite{barch2013function}. Locations of activation for both cases are shown on a standard brain template in Figure~\ref{result2}
\begin{figure}[h]
\begin{center}
\includegraphics[height=2.5cm,width=8cm]{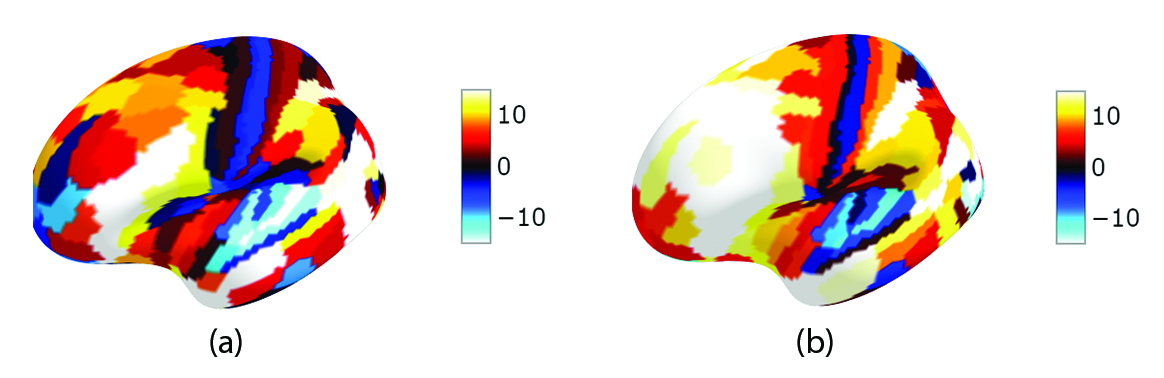}
\end{center}
\caption{Activations in Case A and Case B (lateral view)} 
\label{result2}
\end{figure}

\item
We observed more clear contrasts in activations when the Resp\_R-R was considered as control (Case B) as compared to the RS condition (Case C) (Figure~\ref{result3}).

\begin{figure}[h!]
\begin{center}
\includegraphics[height=2.5cm,width=8cm]{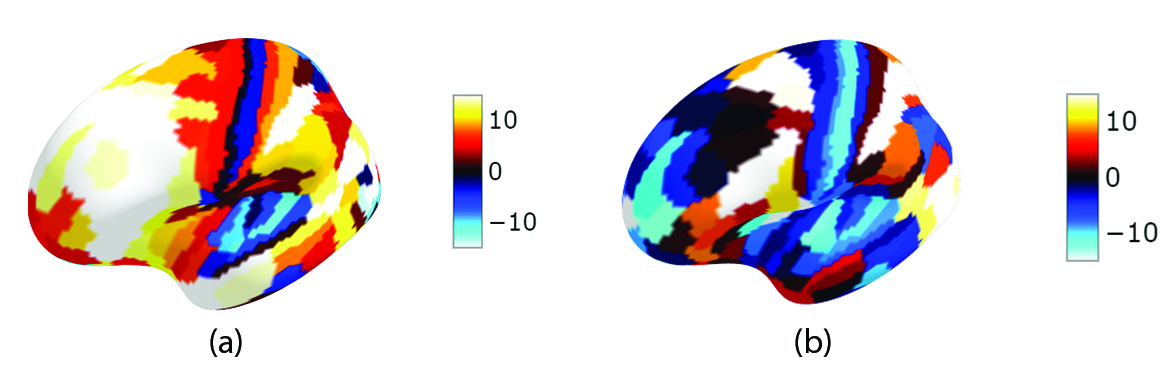}
\end{center}
\caption{Activations in Case B and Case C (lateral view)} 
\label{result3}
\end{figure}

\item
We also observed robust activation in a region close to the pSTS in Case D (Figure~\ref{result4}), which shows contrasts in the activity based on participants' ability to mentalize the interacting shapes. 

\end{enumerate}

\begin{figure}[H]
\begin{center}
\includegraphics[height=4cm,width=7cm]{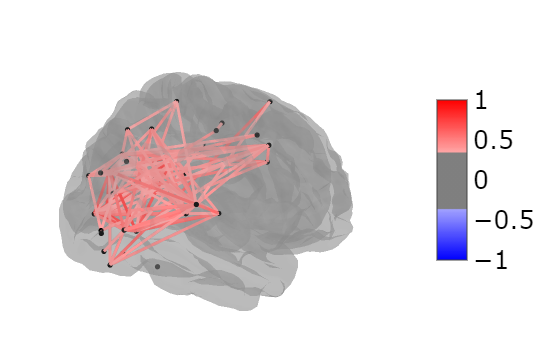}
\end{center}
\caption{Connections between nodes based on the correlation values (edge threshold=0.8, lateral view)} 
\label{connect}
\end{figure}

\begin{figure*}[h]
\begin{center}
  \includegraphics[height=8cm]{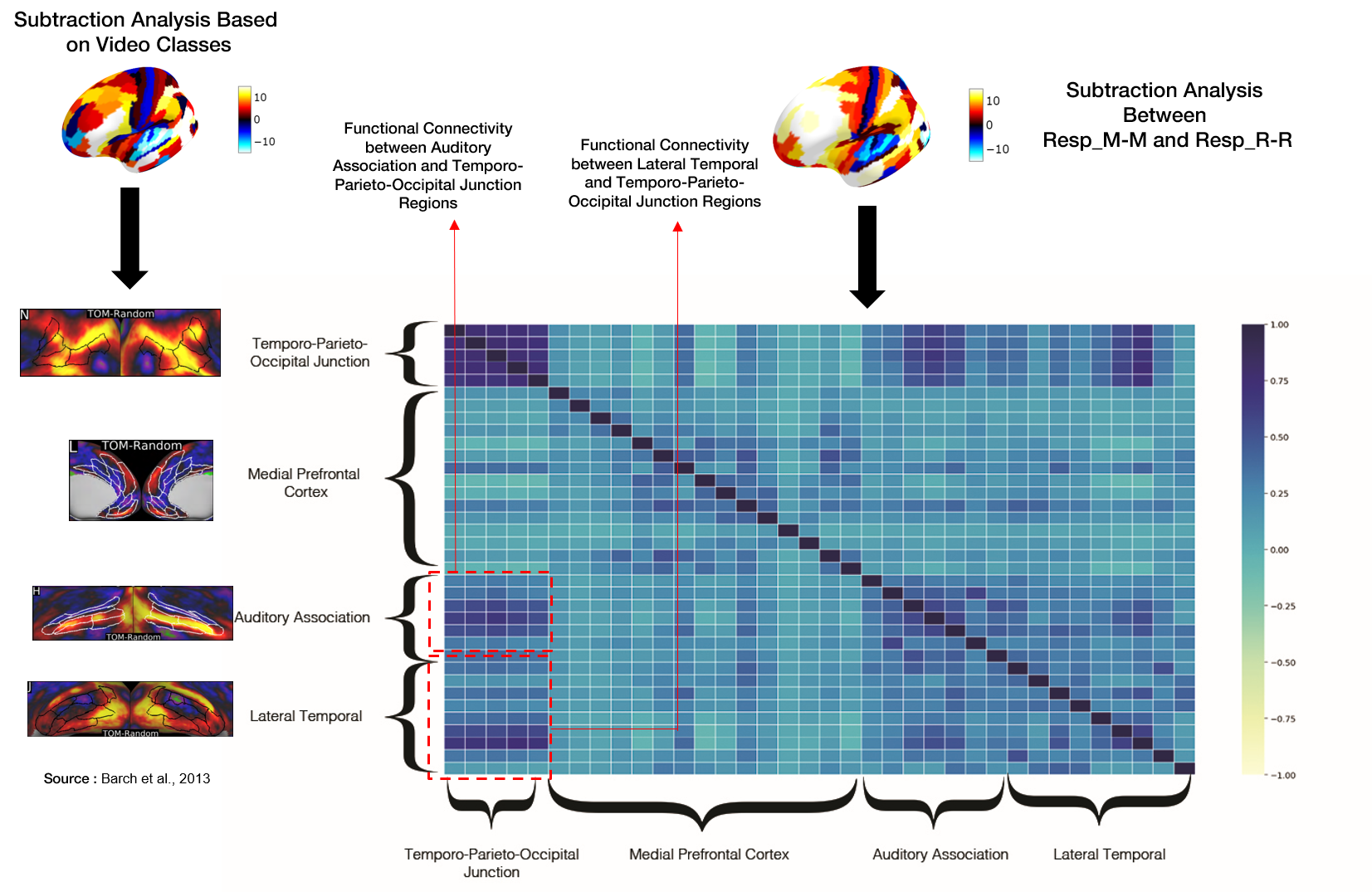}
  \end{center}
  \caption{Correlational analysis for functional connectivity}
  \label{func}
\end{figure*}

\subsection{Functional Connectivity}
Pearson's correlation was computed among the fMRI BOLD activity of ROIs. Strong positive correlations were observed between TPOj and LTC [r = 0.911; p $<$ 0.001] and TPOj and AA [p = 0.876; p $<$ 0.001]. TPOj and mPFC were observed to be weakly negatively correlated [r = -0.167; p = 0.03] (Figure~\ref{func}). Further, the correlation values were thresholded to produce the connectome (Figure~\ref{connect}).

\begin{figure}
\begin{center}
  \includegraphics[width=7cm]{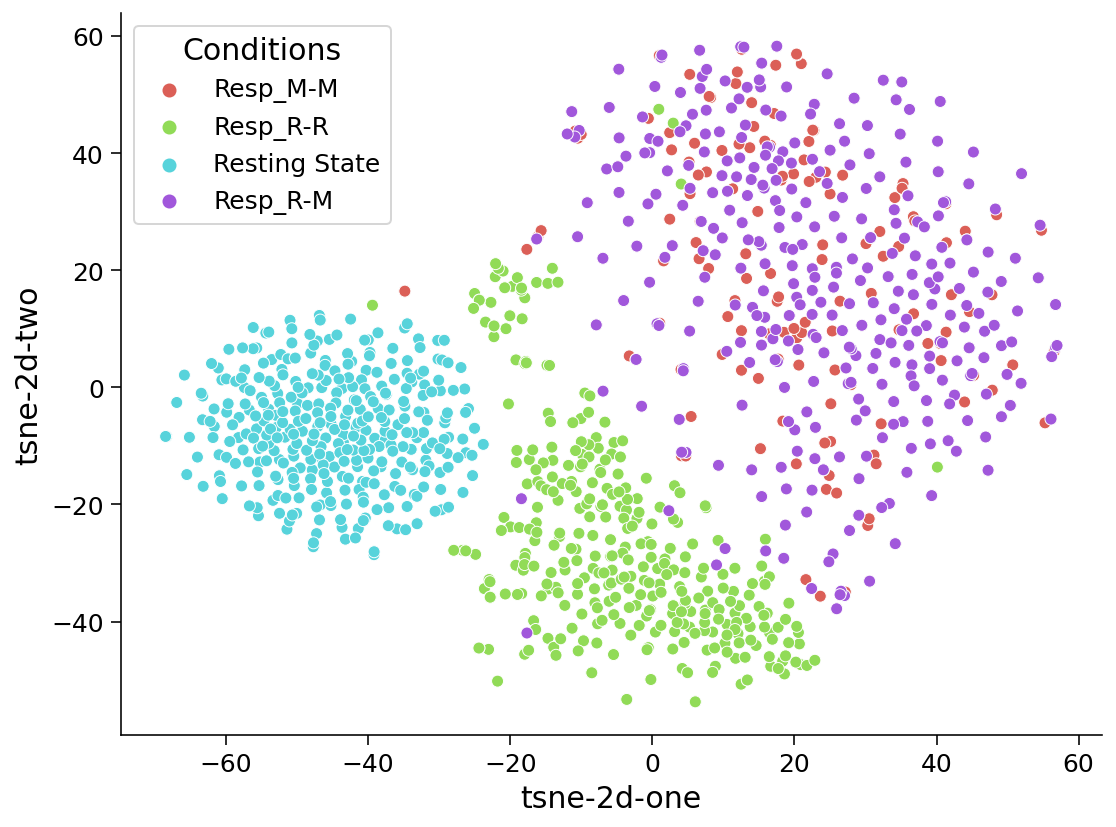}
  \end{center}
  \caption{t-SNE projections for the fMRI data in all the four conditions}
  \label{tsne_full}
\end{figure}

\begin{figure}
\begin{center}
  \includegraphics[width=7cm]{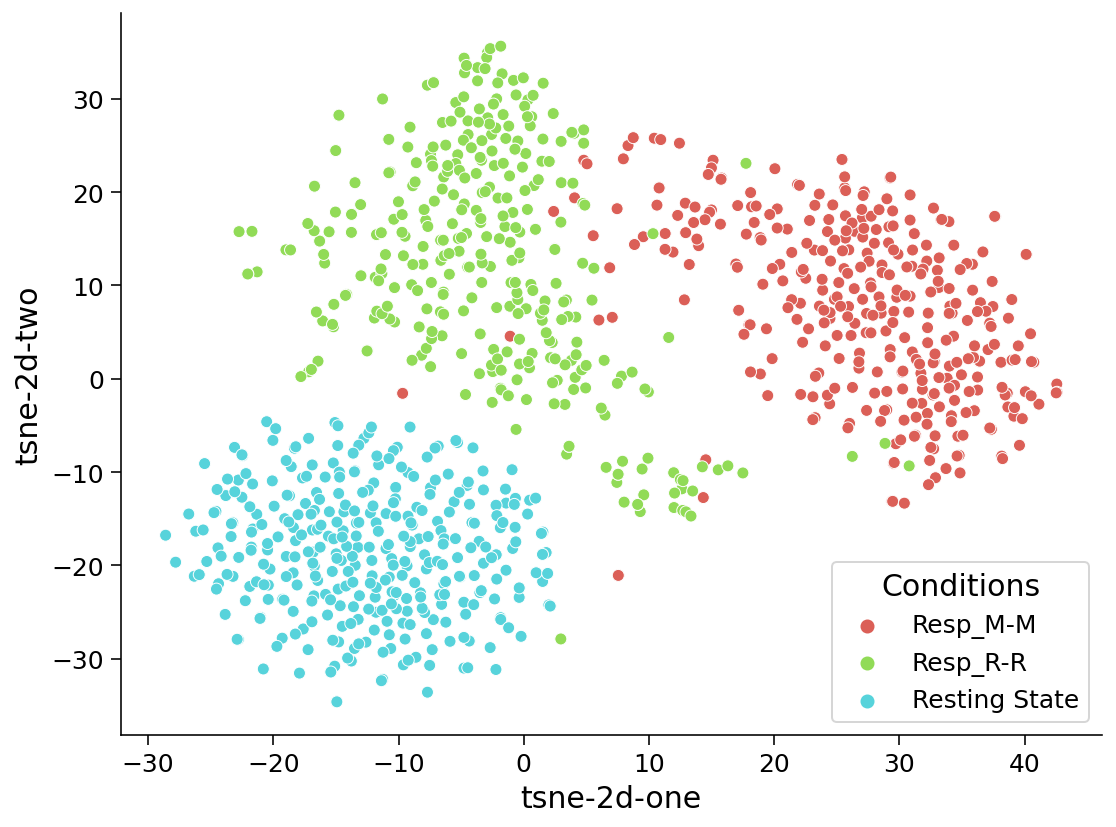}
  \end{center}
  \caption{t-SNE projections for the fMRI data in three selected conditions}
  \label{tsne_three}
\end{figure}

% Three other regions were observed having high activations, which were excluded from our analysis as they are dedicated for cognitive tasks other than empathy. Those regions included Inferior Frontal Cortex, which is responsible for attention and working memory, Medial Temporal Cortex, responsible for spatial memory and long term memory, MT+ Complex and neighboring visual areas which are responsible for visual processing.

\subsection{Dimensionality Reduction}
We used t-Distributed Stochastic Neighbor Embedding (t-SNE) to map the fMRI BOLD time-series data for all the four conditions in two dimensions (Figure~\ref{tsne_full}). However, we observe here that the projections for Resp\_M-M and Resp\_R-M conditions seem to be overlapping. Therefore, for a meaningful interpretation and visualization, we perform another mapping which includes the Resp\_M-M condition, but excludes the Resp\_R-M condition (Figure~\ref{tsne_three}). We suspect that the reason behind the overlap could be the similar condition of shapes that are perceived by participants, but which might elicit incorrect responses due to effects of other activated networks or regions. However, an analysis needs to be performed to explain this observation conclusively. 
In the t-SNE projections corresponding to the three selected conditions, we observe three distinct clusters that could be identified as belonging to the three respective conditions.

\subsection{Predicting Mentalization}
We subsequently used k-nearest neighbour classifier to investigate if these two-dimensional projections of the brain activity could be modelled to predict whether or not a person is mentalizing.

\begin{figure}
\begin{center}
\includegraphics[width=7cm]{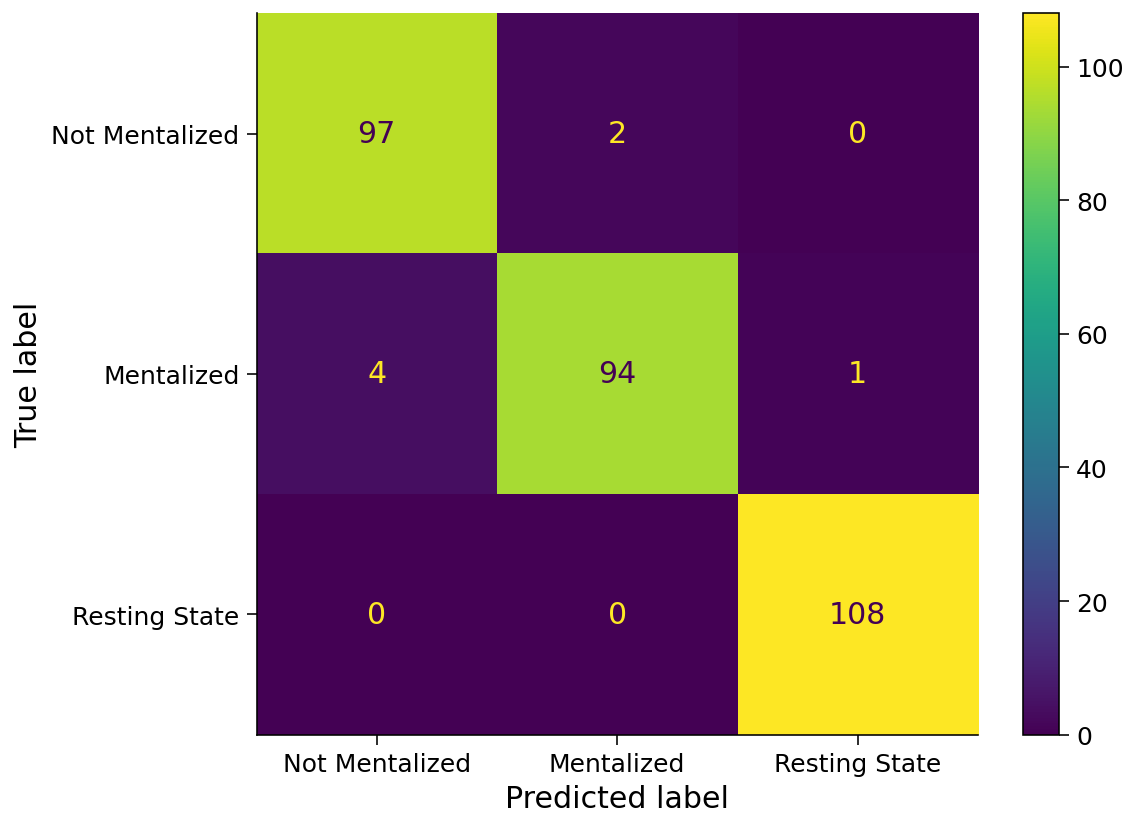}
\end{center}
\caption{Confusion Matrix} 
\label{conf}
\end{figure}

Classification metrics were acquired by evaluating the model on the testing set consisting of brain activity from 30\% (306) of the concatenated fMRI BOLD time-series data (from the four ROIs as discussed in \nameref{roi} section) corresponding to the three selected conditions (Resp\_M-M, Resp\_R-R and Resting State). 
Our model was 97.71\% accurate in making predictions about mentalization on the testing set, as shown along with other metrics in Table~\ref{metric0}. The confusion matrix can be observed in Figure~\ref{conf}.

\begin{table}[h]
\caption{Classification metrics} 
\label{metric0} 
\begin{tabular}{|lllll|}
\hline
\multicolumn{1}{|l|}{}               & \multicolumn{1}{l|}{Precision} & \multicolumn{1}{l|}{Recall} & \multicolumn{1}{l|}{f1-score} & Support \\ \hline
%(A)                                  &                                &                             &                               &         \\ \hline
\multicolumn{1}{|l|}{Not Mentalized}     & \multicolumn{1}{l|}{0.96}      & \multicolumn{1}{l|}{0.98}   & \multicolumn{1}{l|}{0.97}     & 99     \\
\multicolumn{1}{|l|}{Mentalized} & \multicolumn{1}{l|}{0.98}      & \multicolumn{1}{l|}{0.95}   & \multicolumn{1}{l|}{0.96}     & 99     \\ 
\multicolumn{1}{|l|}{Resting State} & \multicolumn{1}{l|}{0.99}      & \multicolumn{1}{l|}{1.00}   & \multicolumn{1}{l|}{1.00}     & 108      \\ \hline
\multicolumn{1}{|l|}{Accuracy}       & \multicolumn{1}{l|}{}          & \multicolumn{1}{l|}{}       & \multicolumn{1}{l|}{0.98}     & 306     \\
\multicolumn{1}{|l|}{Macro avg}      & \multicolumn{1}{l|}{0.98}      & \multicolumn{1}{l|}{0.98}   & \multicolumn{1}{l|}{0.98}     & 306     \\
\multicolumn{1}{|l|}{Weighted avg}   & \multicolumn{1}{l|}{0.98}      & \multicolumn{1}{l|}{0.98}   & \multicolumn{1}{l|}{0.98}     & 306     \\ \hline

% (B)                                  &                                &                             &                               &         \\ \hline
% \multicolumn{1}{|l|}{Mentalized}     & \multicolumn{1}{l|}{0.98}      & \multicolumn{1}{l|}{0.99}   & \multicolumn{1}{l|}{0.99}     & 111     \\
% \multicolumn{1}{|l|}{Not Mentalized} & \multicolumn{1}{l|}{0.99}      & \multicolumn{1}{l|}{0.98}   & \multicolumn{1}{l|}{0.98}     & 93      \\ \hline
% \multicolumn{1}{|l|}{Accuracy}       & \multicolumn{1}{l|}{}          & \multicolumn{1}{l|}{}       & \multicolumn{1}{l|}{0.99}     & 204     \\
% \multicolumn{1}{|l|}{Macro avg}      & \multicolumn{1}{l|}{0.99}      & \multicolumn{1}{l|}{0.98}   & \multicolumn{1}{l|}{0.99}     & 204     \\
% \multicolumn{1}{|l|}{Weighted avg}   & \multicolumn{1}{l|}{0.99}      & \multicolumn{1}{l|}{0.99}   & \multicolumn{1}{l|}{0.99}     & 204     \\ \hline

\end{tabular}
\end{table}

\begin{figure}
\begin{center}
\includegraphics[width=7cm]{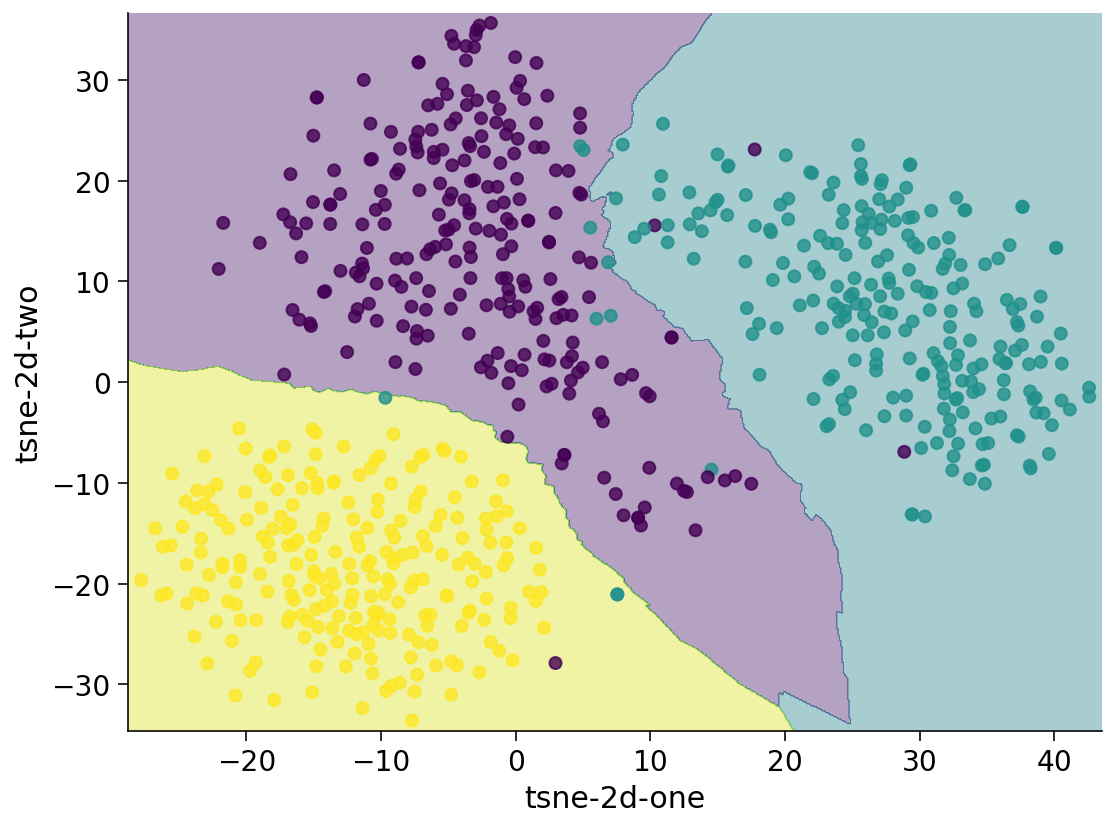}
\end{center}
\caption{Decision Boundaries learnt from the train set} 
\label{decision_train}
\end{figure}

\begin{figure}
\begin{center}
\includegraphics[width=7cm]{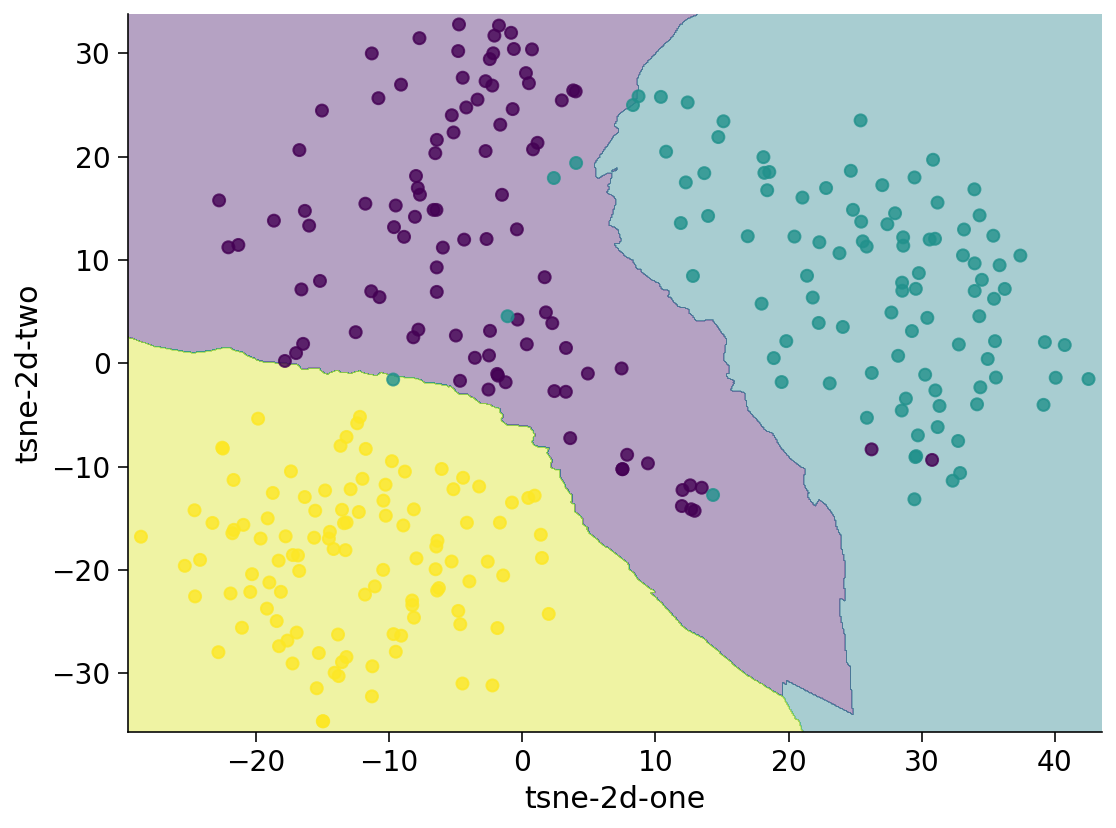}
\end{center}
\caption{Model predictions on the test set along with the decision boundaries} 
\label{decision_test}
\end{figure}

We also plotted and visualised the decision boundaries for our model. Figure~\ref{decision_train} shows the decision boundaries it learnt from the training set and Figure~\ref{decision_test} shows the predictions on the test set along with the decision boundaries.

On the basis of these results, we can conclude that our model is able to make predictions about mentalization with a significant accuracy across participants.

\section{Discussion}

\begin{enumerate}

\item 
We performed subtraction analyses on the fMRI activity to identify brain regions and networks activated while mentalizing animated interacting shapes. We observed that TPOj, that is associated with empathy, showed robust activation during the task. On performing functional connectivity analysis by computing correlations, we established that TPOj is functionally connected with AA and LTC, while mentalizing. In their study, \citeA{silani2013right} explain empathizing as looking at other's world through our emotions and mentions, "When assessing the world around us and our fellow humans, we use ourselves as a yardstick and tend to project our own emotional state onto others." They observed the neurological correlates for empathy by varying the activity in temporo-parieto-occipital junction (TPOj) \cite{silani2013right}. As TPOj is associated with empathy, we speculate mentalizing and empathy may co-occur even in this case, as previous research posits \cite{cerniglia2019intersections, schnell2011functional, hoffmann2016preserved}. This finding suggests that humans might be able to empathize with such kind of interacting shapes. However, further behavioural investigations by introducing a questionnaire based on empathy in the task is required to conclude this. Since the previous studies have been limited to human features, including cues like facial expressions, body posture etc., through which people are able to mentalize and empathize, whether empathy can be evoked through exposure to simple, non-verbal stimuli that elicit mental state attributions i.e. mentalizing, in spite of lacking human feature cues, is an interesting question for future work.  

\item 
In the study done by \citeA{barch2013function}, activations in the mPFC were not observed to be robust (as replicated by us in Figure~\ref{result2}a), although previous studies show otherwise. The mPFC has been found to play an important role in understanding other's and one's own nature \cite{cerniglia2019intersections}. However, by considering paricipants' responses, we observed significant activation in the mPFC for the mentalizing task (Figure~\ref{result2}b). Hence, our study explained the lack of robust activation in the same region in previous research that had not considered the responses. \citeA{barch2013function}.
% By taking into account participants' responses, we obtained robust activations in mPFC, that plays an important role in understanding other’s and one’s own nature, plays an important role in mentalizing. Hence, our study explained the lack of activation in the same region in previous research that had not considered the responses.

\item 
We noted that for our study, Resp\_R-R is a better control than Resting State (RS). This is because, the former controls for the cognitive feature to be studying (i.e. mentalizing), while the later is simply an absence of the task stimuli. Additionally, RS has been found to be associated with an individual's mentalizing ability \cite{hoffmann2016preserved}. Hence, for our task, absence of the task stimuli (i.e. mentalizing in RS) is not a good control condition as compared to the condition which differs only in the cognitive feature to be studied (i.e. mentalizing). This reiterates the finding that RS might not always act as a good control condition \cite{mastrovito2013interactions}, and that control conditions should be task-specific. 

%  This is because the former controls for the cognitive feature to be studying (i.e. mentalizing), while the later is simply an absence of the task stimuli. 
%  These results reaffirm that control conditions should be task-specific.

\item 
In Case D, we obtained robust activations in a region close to the pSTS, in which the contrasts depict participants' 'ability to mentalize'. In clinical studies, this region has been found to be less activated in autistic individuals \cite{castelli2002autism, alaerts2014underconnectivity}. In previous studies, impairment in mentalization and empathy have been linked to various psychological disorders such as autism \cite{frith2001mind, castelli2002autism, white2011developing, abell2000triangles}, psychopathy \cite{decety2013fmri} and schizophrenia \cite{russell2006you}. Hence, we believe an analysis similar to ours could be used for research in disorders linked with difficulties in mentalization.  

\item 
We used t-Distributed Stochastic Neighbor Embedding (t-SNE) to project the fMRI BOLD time-series data for selected conditions across two dimensions to visualise patterns within the data. We observed that these projections are distinctively clustered according the respective condition.
In light of this observation, we modelled the prediction of mentalization using a k-nearest neighbor classifier. When trained on the two-dimensional t-SNE mappings of the corresponding brain activity in the four ROIs, the model was able to generate distinct and clear decision boundaries and returned impressive test accuracy in predicting about mentalization. Along with our findings and observations of the brain activity during perception of these animations, we propose our model as an approach towards developing a brain-activity based model to detect ToM (Theory of Mind) difficulties, which could be useful in research about disorders like Autism Spectrum Disorder (ASD) as well as assessment of mentalization-based treatments. This however must be taken with two pinches of salt - one, that this approach still needs to be extended to clinical population \cite{kazeminejad2019topological}; two, that fMRI is an expensive option to use for diagnostic purpose. However, as these videos (as part of the Frith–Happé Animations Test) have been successfully used to measure ToM ability in ASD \cite{white2011developing}, we hope that such an approach could be helpful in corroborating the diagnoses in cases where the test results lack a degree of certainty.

% \includegraphics[width=3in]{Figures/caseB1.png}
% Impairments in cognitive empathy have been found in autistic individuals. . This could help in the assessment of disorders like autism. 

% Mentalizing, empathy and mirroring of emotional response are closely related. Hence, our approach on a neuronal level can help strengthen the foundation of mirror neurons as the brain region `OFC' is observed to be a point of intersection between affective cognition and emotional mirroring \cite{cerniglia2019intersections}. 

% \item
% Empathy is considered important for patient-care, and these analyses could be helpful in devising training programs for students in medical education to provide quality health care. Moreover, by monitoring empathy, even psychopaths can be treated and can help lowering violent crimes \cite{decety2013fmri}. For effective psychotherapy, \citeA{fonagy2012mentalization} posit that mentalization is the core, as it promotes non-judgemental attitude of knowing others through inquiry, instead of making assumptions or formulations.

\end{enumerate}

% The code to reproduce the results could be found at: https://gofile.io/d/2PWOzS

% Research has been done to form bio-markers to detect autism spectrum disorder (ASD) using modelling approaches. 

% \subsection{Footnotes}

% Indicate footnotes with a number\footnote{Sample of the first
% footnote.} in the text. Place the footnotes in 9~point font at the
% bottom of the column on which they appear. Precede the footnote block
% with a horizontal rule.\footnote{Sample of the second footnote.}

% \subsection{Figures}

% All artwork must be very dark for purposes of reproduction and should
% not be hand drawn. Number figures sequentially, placing the figure
% number and caption, in 10~point, after the figure with one line space
% above the caption and one line space below it, as in
% Figure~\ref{sample-figure}. If necessary, leave extra white space at
% the bottom of the page to avoid splitting the figure and figure
% caption. You may float figures to the top or bottom of a column, and
% you may set wide figures across both columns.

% \begin{figure}[H]
% \begin{center}
% \fbox{CoGNiTiVe ScIeNcE}
% \end{center}
% \caption{This is a figure.} 
% \label{sample-figure}
% \end{figure}

\section{Acknowledgments} 

We thank Dr. Laura Mikula (York University) and Dr. Marlene Cohen (University of Chicago), for their valuable inputs and useful discussions.

\bibliographystyle{apacite}

\setlength{\bibleftmargin}{.125in}
\setlength{\bibindent}{-\bibleftmargin}

\bibliography{CogSci_Template}

\end{document}